\documentclass[reprint,amsmath,amssymb,aps,twocolumn]{revtex4-2}

\usepackage{graphicx}
\usepackage{bm, slashed}
\usepackage{tikz} 
\usepackage{hyperref}
\usepackage{comment}
\usepackage{mathdots,bbold}
\usepackage{soul}
\usepackage{braket}
\usepackage[normalem]{ulem}
\setcitestyle{square,numbers}
\usepackage{pgfplotstable}
\usepackage{mathtools}
\usepackage{enumerate}

\begin{document}
	
	\title{Renormalized dual basis for scalable simulations of 2+1D compact quantum electrodynamics}
	
	\author{Marc Miranda-Riaza}
	\email{marc.miranda.riaza@uab.cat} 
	\affiliation{
		Departament de Física, Universitat Autònoma de Barcelona, 08193 Bellaterra, Spain
	}
	
	\author{Pierpaolo Fontana}
	\email{pierpaolo.fontana@uab.cat}
	\affiliation{
		Departament de Física, Universitat Autònoma de Barcelona, 08193 Bellaterra, Spain
	}
	
	\author{Alessio Celi}
    \email{alessio.celi@uab.cat}
	\affiliation{
		Departament de Física, Universitat Autònoma de Barcelona, 08193 Bellaterra, Spain
	}
	
	\date{\today}
	
	\begin{abstract}
		The classical and quantum simulation of lattice gauge theories (LGTs) with Lie groups is hindered by the infinite-dimensional Hilbert space of gauge degrees of freedom. In a recent work [Phys. Rev. X {\bf 15}, 031065 (2025)], we introduced a new truncation scheme -- here renamed as {\it Renormalized Dual Basis} (RDB) -- based on the resolution of the single-plaquette problem, and demonstrated its performance for $\text{SU}(2)$ LGTs. In this paper, we apply the RDB to compact quantum electrodynamics (cQED) in three spacetime dimensions (2+1D). We variationally determine the ground state of the theory for small lattices with periodic (for pure gauge) and open (in presence of fermionic matter) boundary conditions, achieving improved precision for the plaquette operator compared to previous approaches. By leveraging tensor networks, we extend the study to larger lattices and demonstrate the scalability of the method. Overall, we show that the RDB provides an efficient description across all coupling regimes.
	\end{abstract}
	
	\maketitle
	
	\section{Introduction}
    Lattice gauge theories (LGTs) provide a non-perturbative framework for exploring both fundamental interactions in high-energy physics and emergent gauge phenomena in condensed matter systems \cite{Wilson1974,Peskin,Scwhartz,Wen,Savary2017}. In recent years, the Hamiltonian formulation of LGTs \cite{Kogut-Susskind,Creutz1985,Ligterink2000} has gained renewed attention \cite{ZoharPRL2012,Banerjee2012,tagliacozzo_NC2013,Banuls2013-rb,PRXTagliacozzoCeli2014,RicoPRL2014,zohar2015quantum,Wiese,Dalmonte-Montangero,PichlerPRX2016,Preskill2018Nov,barros2020gauge,banulsRPP2020,Banuls2020,aidelsburger2022,davoudi_QSforHEP2022,Halimeh2023Oct,DiMeglio2023Jul,GaugedGaussian2024,osborne22}, as Hamiltonian calculations performed on quantum computers or with quantum-inspired algorithms promise to circumvent the sign problem that affects Euclidean path integral Monte Carlo methods \cite{Muroya03,troyer_wiese2005,gattringer2009,fukushima2010,Delia2019,Nagata22}. The first experimental results of gauge theories in digital and analog computers in two spacetime dimensions \cite{Martinez2016,KlcoPRA2018,Schweizer2019,Yang2019,Kokail2019,Mil2020,SuracePRX2020,KlcoPRD2020,NguyenPRXQuantum2022,ZhaoScience2022,Mildenberger2022,ChiralBF2022,SuPRR2023,Zhang2023preprint, ObservationStringBreaking2024, Schuster2024, than2024phasediagramquantumchromodynamics, Melzer2025, Liu25}, and proof-of-principle simulations in higher dimensions with $\mathbb{Z}_2$ \cite{Semeghini2021,Satzinger2021} and $\text{U}(1)$ \cite{Meth2023Oct, GonzalezCuadra2025, crippa2024analysisconfinementstring2} groups have fueled these expectations. However, phenomenologically relevant gauge theories display infinite-dimensional Hilbert spaces that hinder large-scale computations on quantum computers or with quantum-inspired methods. The viability of these computations depends on whether one can efficiently truncate the gauge degrees of freedom to a finite number of states  across all lattice sizes and coupling strengths, while faithfully preserving the underlying gauge structure. Recently, we provided a solution by proposing an efficient formulation for non-Abelian LGTs and performed a proof-of-principle computation on a minimal torus \cite{Fontana2024}. Here, we show that this method significantly improves on previous algorithms and enables large-scale simulations, making it a strong candidate for advancing both near- and long-term quantum(-inspired) computations. To emphasize the content of the method, we rename it here to Renormalized Dual Basis (RDB). 
    
    The quest for efficient methods to deal with infinite dimensional gauge degrees of freedom has become central in the Hamiltonian approach to LGTs. The schemes explored so far include discretizing the gauge group manifold, by reducing it to finite subgroups \cite{KuhnPRA2014,notarnicola_JPA2015,ErcolessiPRD2018,Haase2021resourceefficient,PaulsonPRX2021,DanielGCPRL2022,DiscreteLammPRD2022,Zache2023fermionquditquantum,PradhanPRD2023,BauerPRD2023,Bauer2023,BalliniPRD2024} or by finite partitionings \cite{Hartung2022Mar,Jakobs:2025rvz}, defining quantum link models and electric field-like truncations \cite{HORN1981,Chandrasekharan1997,Wiese,quantumSpinIceReview}, using improved Hamiltonians \cite{LammPRL2022}, deforming the algebra through quantum groups \cite{ZachePRL2023} and parametrizing gauge-invariant states with neural networks \cite{Favoni2022,luo2022gaugeequivariantneuralnetworks,spriggs2025accurategroundstatessu2}. All these methods present shortcomings. Truncations based on the irreducible representations of the gauge group are bound to fail in the weak coupling regime, or require additional dimensions \cite{Chandrasekharan1997}, while deformations of the algebra come at the cost of not ensuring the variational nature of the obtained states. A successful approach should scale efficiently with the system size, accurately accommodate all necessary terms in the Hamiltonian, and remain efficient across the full parameter space of the theory.

    The method that we recently introduced is designed to overcome these limitations \cite{Fontana2024}. In the Kogut--Susskind (KS) formulation of Hamiltonian LGTs the presence of the Gauss law implies that the physical Hilbert space does not decompose as a simple tensor product of the gauge degrees of freedom. This fact converts basis optimization into a nonlocal problem. Through a dualization procedure \cite{Kaplan-Stryker, MathurPRD2015} that ensures gauge invariance we are able to turn basis optimization into a local task. After dualizing the theory in terms of loop variables, we truncate the gauge degrees of freedom through the resolution of a single-plaquette eigenvalue problem with adjustable coupling. By determining the optimal single-plaquette basis variationally, we incorporate the renormalization of the loop basis due to the coupling between the loops. For this reason, we name here this method Renormalized Dual Basis. It allows for efficient classical precomputation of the basis, significantly reducing the computational cost of subsequent simulations of LGTs on both classical and quantum platforms for any value of the coupling. Its application to $\text{SU}(2)$ LGT on a minimal torus shows percent precision in the computation of the plaquette operator for any coupling with just 5 states per plaquette \cite{Fontana2024}, a number compatible with current quantum hardware used in trapped-ions platforms \cite{Meth2023Oct}. In this paper, we apply the RDB to $\text{U}(1)$ LGTs and evaluate its performance. For the pure gauge theory on small lattices, we benchmark the performance of the RDB against existing formulations, showing improved accuracy and resource efficiency. We then extend this analysis by including matter fields in a minimal compact quantum electrodynamics (cQED) setup. Finally, by leveraging tensor network techniques, we demonstrate the scalability of the RDB for large-scale LGT simulations. 
    
    The paper is organized as follows. In Sec. \ref{RDBAbstract} we review the rational and key steps of the RDB introduced in \cite{Fontana2024} and applied to $\text{SU}(N)$ LGTs. In Sec. \ref{RDBcQED} we present the application of the RDB to $\text{U}(1)$ LGTs. First, we remind the properties of cQED and its resource-efficient formulation in terms of strings and rotators \cite{Haase2021resourceefficient}. Then, we illustrate the identification of the optimal local basis. We present the results of the RDB and evaluate its performance in Sections \ref{basis_performance} and \ref{large_scale_section}. In the former, we compare with previous proposals for resource-efficient $\text{U}(1)$ computations and evaluate the performance of the RDB in a minimal cQED setup. In the latter, we perform computations on larger lattices with tensor network algorithms. In Sec. \ref{conclusions}, we discuss future directions of work. In the Appendices [\ref{PBC_OBC_details}-\ref{local_basis_computations}] we present all the numerical and computational details. 
        \begin{figure}[t!]
		\centering
		\includegraphics[width=1.0\linewidth]{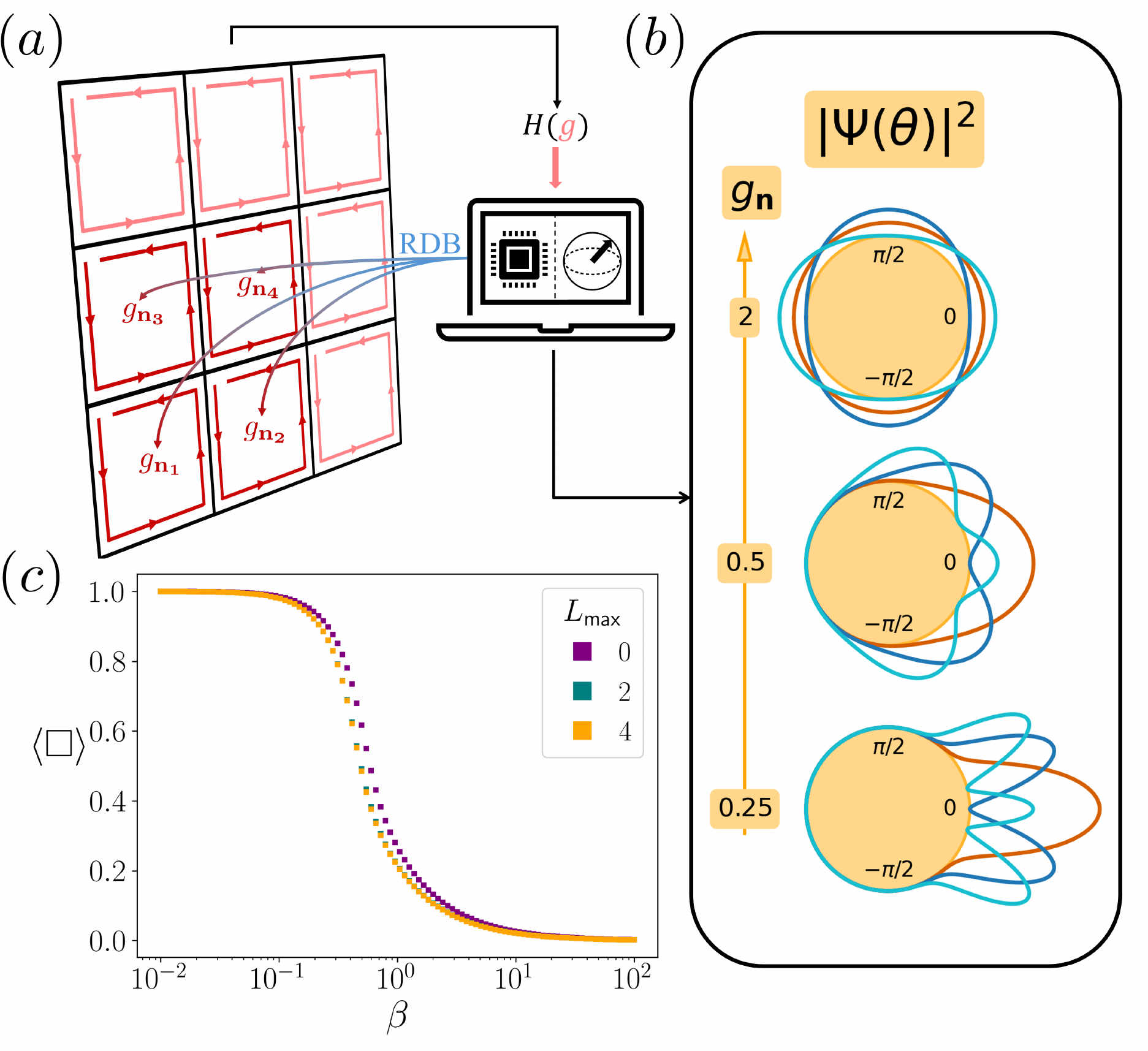}
		\caption{\textbf{Representation of the Renormalized Dual Basis as applied to $\text{U}(1)$ lattice gauge theories}. (a) We get a finite Hamiltonian by truncating the dual Hamiltonian (see Fig. \ref{2d_lattice_vars}), built of plaquettes, in an optimal local basis. The local basis for the $\bf n$ plaquette is obtained from the resolution of a single-plaquette problem for arbitrary basis parameter $g_{\bf n}$ and is variationally optimized by minimizing the energy over $g_{\bf n}$, see Sec. \ref{LocalBasisU1}. (b) Ground (in dark orange), first- (in light blue), and second-excited (in dark blue) states of the single-plaquette problem by varying coupling strength $g_{\bf n}$, from weak $(g_{\bf n}\ll1)$ to strong $(g_{\bf n}\gg1)$ coupling. The single-plaquette eigenbasis can be determined with arbitrary precision for any $g_{\bf n}$. (c) The expectation values of the plaquette operator $\langle\square\rangle$, defined in Eq. \eqref{HB_expval_definition}, are computed by retaining $L_{\text{max}} +1$ states per plaquette and quickly converge to the exact result for any value of the coupling constant $\beta=(2g^2)^{-1}$.}
		\label{RDBConcept}
	\end{figure}
    
	\section{\label{RDBAbstract}Renormalized Dual Basis}
    The choice of an appropriate basis for truncating the Hilbert space of gauge degrees of freedom is a crucial step in the Hamiltonian formulation of LGTs. In this Section we review the RDB that we introduced in \cite{Fontana2024}, the essential steps of which consist in:
    \begin{enumerate}[I)]
        \item A dualization procedure that allows to turn basis optimization into a local task.
        \item The identification and optimization of the local basis in the one-parameter family of eigenbases of the single-plaquette Hamiltonian.
    \end{enumerate}

    \subsection{Dualizing into local basis optimization}
    
    The dualization, see Fig. \ref{RDBConcept}(a), allows us to reformulate the Hamiltonian treatment of LGTs in terms of local degrees of freedom. In the standard link-based representation, the Gauss law ties together extended regions of the lattice and makes basis optimization a global problem. Through the dualization, plaquettes (and large loops in case of periodic boundary conditions) become the natural objects and the Hamiltonian splits into a local part acting on each individual plaquette, i.e., a single-plaquette Hamiltonian \cite{Ligterink2000}, and a nonlocal one that couples different plaquettes. This separation is crucial: it allows us to treat locally the problem of basis selection and the subsequent Hilbert space truncation, while keeping track of the structure of the nonlocal terms. We exploit this separation to establish a systematic many-body treatment of LGTs \cite{Fontana2024}, based on the observation that the space of physical states is the tensor product of plaquette states.

    \subsection{Identification of the optimal local basis}
    
    Once the plaquette is identified as the local object through the duality, the selection of the proper local basis allows us to reduce the resources required in the computation or simulation of LGTs. This identification is unavoidable for LGTs beyond one spatial dimension with continuous gauge groups. Continuous gauge groups imply that the plaquettes, as the links in the original formulation, have an infinite-dimensional Hilbert space, while more spatial dimensions imply non-commuting electric and magnetic terms that prevent the truncation to a common eigenbasis. In fact, since Hamiltonians at different couplings do not commute, there is no coupling-independent basis that is optimal across all the regimes.
        
    Since the single-plaquette Hamiltonian has a discrete spectrum, we use its eigenstates to obtain a basis for any value of the coupling. Notably, the single-plaquette Hamiltonian depends on the coupling strength and, thus, provides a one-parameter family of eigenstates, see Fig. \ref{RDBConcept} (b). A priori, we have the freedom to span each plaquette in a different basis of this family by independently tuning the parameters for each plaquette. We identify the optimal local basis at a given coupling by variationally minimizing the ground-state energy over those parameters, see Fig. \ref{RDBConcept}(a,b). In practice, the basis of each plaquette is truncated to a finite number of $L_{\text{max}}+1$ states and the minimization is repeated for different truncations. The exact result is retrieved for $L_{\text{max}}+1\rightarrow\infty$. Combined with variational optimization, the one-parameter family of bases can interpolate between weak and strong coupling and adapt to the presence of different terms in the Hamiltonian, i.e., matter and nonlocal gauge terms, see Fig. \ref{RDBConcept}(c). The variational optimization is to be understood as a renormalization of the local basis that is induced by the nonlocal Hamiltonian terms and partly absorbs its effects. 
    
    The solution of the single-plaquette problem, as well as the required matrix elements needed for further use in LGT computations, can be obtained very efficiently on a classical computer. Moreover, by avoiding discretization in parameter space (see efforts in this direction in \cite{Bauer2023} for $\text{SU}(2)$ and \cite{Haase2021resourceefficient,BauerPRD2023} for $\text{U}(1)$, to which we compare below) and by precomputing classically the single-plaquette basis, we are able to significantly reduce the requirements for further classical and, potentially, quantum computations. Also, the single-plaquette Hamiltonian readily incorporates the magnetic interaction and avoids the huge increase in local dimension that other local basis optimization techniques face when approaching the weak coupling limit \cite{majcen2025optimallocalbasistruncation}.
    
    \section{\label{RDBcQED} Renormalized Dual Basis for compact Quantum Electrodynamics}
    
    As explained in the previous Section, the dualization and the identification of the optimal local basis through the single-plaquette Hamiltonian are the key ingredients of the RDB. Since the dualization and the single-plaquette Hamiltonian in $\text{U}(1)$ LGT are simpler than in the $\text{SU}(N)$ LGT discussed in \cite{Fontana2024}, we take the opportunity to illustrate the RDB method in a pedagogical manner.
    \begin{figure}[t!]
		\centering
		\includegraphics[width=1.0\linewidth]{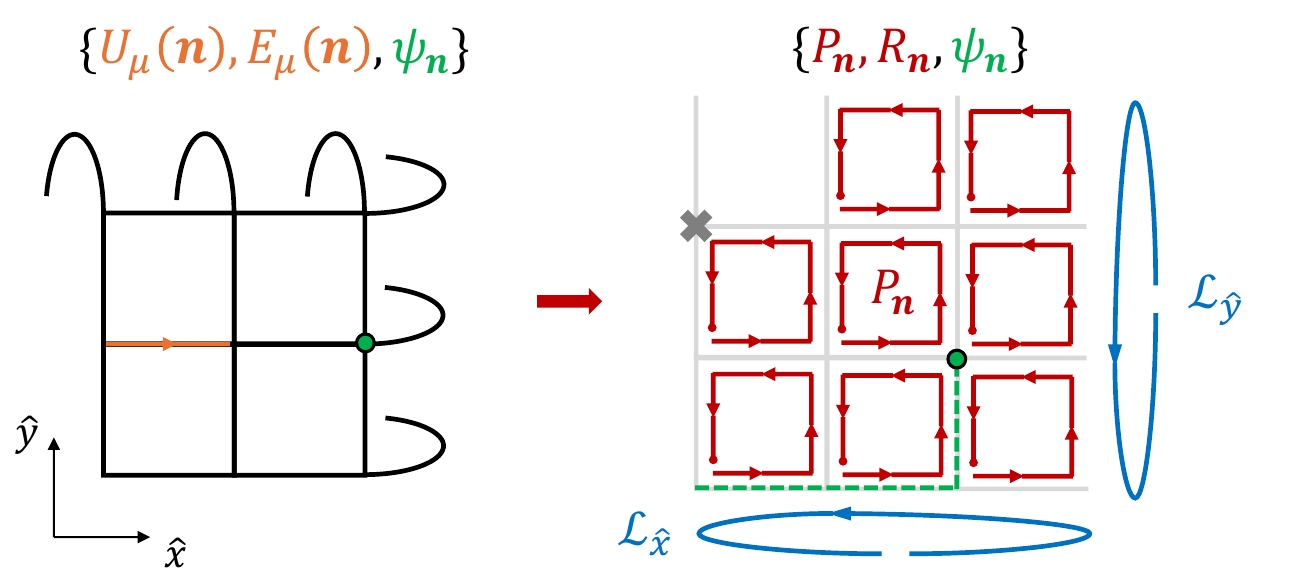}
		\caption{Representation of the fundamental gauge variables in the Kogut--Susskind formulation (left lattice), the links $U_\mu({\bf n})$ and the electric fields $E_\mu({\bf n})$, and the dual variables (right lattice), the plaquettes $P_{{\bf n}}$ and the rotators $R_{{\bf n}}$. In the case of periodic boundary conditions, we have to consider the large loops wrapping the lattice, i.e., $\mathcal{L}_{\hat{x},\hat{y}}$, and the absence of the plaquette in the reference site ${\bf n}_R\equiv(0,N-1)$ (dark grey cross). Dynamical fermions (staggered formulation) are replaced in the dual formulation by electric strings connecting the origin to the other sites \cite{Haase2021resourceefficient}.}
		\label{2d_lattice_vars}
	\end{figure}
    
    \subsection{\label{DualLBO} Dualizing into local basis optimization}
    We consider the Hamiltonian formulation of 2+1D cQED on a rectangular lattice $\Lambda$ of size $N_x\times N_y$, both with open and periodic boundary conditions. We denote each lattice site by ${\bf n}=(n_x,n_y)$, with $n_{\mu}\in\{0,\ldots,N-1\}$ and unit vectors $\mu=\hat{x},\hat{y}$. According to the KS formulation \cite{Kogut-Susskind}, the lattice links, or Wilson operators, $U_\mu({\bf n})$ encode the gauge potentials $A_\mu$ as $U_\mu({\bf n})\sim \exp{(i\,a\,g\, A_\mu({\bf n}))}$, where $a$ and $g$ are respectively the lattice spacing and the bare coupling, which is proportional to the bare electric charge. The Wilson operators are canonically conjugate to the electric fields $E_\mu({\bf n})$, satisfying the commutation relations
	\begin{equation}
		[E_\mu({\bf n}),U_\nu({\bf m})]=-\delta_{\bf{nm}}\delta_{\mu\nu} U_\mu(\bf{n}),
		\label{standard_CR}
	\end{equation} 
	where $\delta_{ij}$ is the Kronecker delta. The pure gauge part of the Hamiltonian is
	\begin{align}
		\nonumber
		H_G&=H_E+H_B\\
		&=\frac{g^2}{2}\sum_{\mu,{\bf n}}E^2_\mu({\bf n})+\frac{1}{2g^2a^2}\sum_{{\bf n}}[2-U_{\hat{x}\hat y}({\bf n})-U^\dagger_{\hat x\hat y}({\bf n})],
		\label{HG_definition}
	\end{align}
	where we defined the plaquette operator $U_{\hat x\hat y}({\bf n})\equiv U_{\hat x}({\bf n})U_{\hat y}({\bf n}+\hat x)U^\dagger_{\hat x}({\bf n}+\hat y)U^\dagger_{\hat y}({\bf n})$. By sending the lattice spacing $a\to 0$, Eq. \eqref{HG_definition} approaches the Hamiltonian of the electromagnetic field in the continuum. In the following we set $a=1$ without loss of generality.
	
	  For the matter fields, we consider Dirac fermions in the staggered representation, with fermion field $\psi_{\bf n}$ destroying (creating) a particle (antiparticle) on even (odd) sites of the lattice $\Lambda$ \cite{Kogut-Susskind,Rothe,Creutz1985}. They satisfy canonical anti-commutation relations, $\{\psi^\dagger_{{\bf n}},\psi_{{\bf m}}\}=\delta_{\bf{nm}}$, and their Hamiltonian contributions are
	\begin{align}
    \nonumber
    H_M &= H_m+H_k\\
        &= m\sum_{{\bf n}}(-1)^{n_x+n_y}\psi^\dagger_{{\bf n}}\psi_{{\bf n}}\\
        \nonumber
        &+ \kappa\sum_{{\bf n}}\Big(i\psi^\dagger_{{\bf n}}U_{\hat x}({\bf n})\psi_{{\bf n}+\hat{x}} \\
        &\phantom{+ \kappa\sum_{{\bf n}}\Big(}
          + (-1)^{n_x+n_y+1}\psi^\dagger_{{\bf n}}U_{\hat y}({\bf n})\psi_{{\bf n}+\hat{y}}
          + \text{H.c.}\Big).
    \label{HM_definition}
    \end{align}
	The first contribution represents the mass term of the staggered fermions, acting as a staggered chemical potential, while the second contribution is the kinetic term, describing tunneling processes for the generation or annihilation of particle-antiparticle pairs, along with the relative change of the electric field string connecting them, needed to ensure a gauge-invariant Hamiltonian term.
   
	Gauge symmetry at the Hamiltonian level requires that physical states satisfy the Gauss law, which on the lattice reads
	\begin{equation}
		\mathcal{G}({\bf n})|\psi_{\text{phys}}\rangle=0,\qquad\forall\;{\bf n}\in\Lambda,
		\label{Gauss_law}
	\end{equation}
    where $\mathcal{G}({\bf n})=\sum_{\mu=\hat{x},\hat{y}}[E_\mu({\bf n})-E_\mu({\bf n}-\mu)]-q_{{\bf n}} - Q_{\bf n}$ are the generators of local $\text{U}(1)$ transformations, $q_{{\bf n}}\equiv\psi^\dagger_{{\bf n}}\psi_{{\bf n}}-\frac{1-(-1)^{n_x+n_y}}{2}$ is the fermion charge at the site $\bf n$, and $Q_{\bf n}$ represents the static charge background. In this work, we do not consider static charges, thus $Q_{\bf{n}}=0\;\forall \mathbf{n} \in\Lambda$. The condition in Eq. \eqref{Gauss_law} selects the physical states out of the whole Hilbert space of the Hamiltonian, $\mathcal{H}_{\text{phys}}\subset\mathcal{H}$. As emphasized above, the Gauss law couples the link degrees of freedom and turn the basis optimization problem into a nonlocal one. In other words, the gauge-invariant Hilbert space $\mathcal{H}_{\text{phys}}$ does not factorize into a simple tensor product of the KS link degrees of freedom.

    To recast the problem of basis selection as a local optimization task, we consider the dual formulation of the theory, obtained by applying a sequence of canonical transformations to the original KS links and conjugate electric fields \cite{MathurPRD2015}. We recently presented the dualization procedure for $\text{SU}(N)$ LGTs with periodic boundary conditions \cite{Fontana2024}, complementing earlier results for open boundary conditions \cite{MathurPRD2015}. The application of this procedure to $\text{U}(1)$ LGTs yields the dual formulation presented in \cite{Haase2021resourceefficient,PaulsonPRX2021, Bender-Zohar}, to which we refer the reader for a detailed exposition. In the present work, we adopt these results and provide only a minimal description.
		
	To derive the dual Hamiltonian in terms of physical variables, we resolve the gauge symmetry at each lattice site to get rid of redundant degrees of freedom and obtain an unconstrained Hamiltonian.  In one spatial dimension this procedure allows for a complete elimination of the gauge field in terms of matter fields, leading to an effective Hamiltonian with long-range interactions \cite{HamerPRD1997,MuschikNJP2017}, expected to arise when solving the local constraints. In higher spatial dimensions, the gauge field is physical too, and its complete elimination is in general not possible \footnote{For the special case of topological gauge theories like Chern-Simons, see \cite{ChisholmEncoding}, for instance}. In terms of fundamental lattice variables, this procedure realizes the transformation \cite{Drell1879}
	\begin{equation}
		\{U_\mu({\bf n}),E_\mu({\bf n})\}\quad\rightarrow\quad\{P_{{\bf n}},R_{{\bf n}}\}
		\label{DOF_transformation}
	\end{equation}
	on the initial KS variables. The transformation is represented in Fig. \ref{2d_lattice_vars}. The dual degrees of freedom on the right-hand side of Eq. \eqref{DOF_transformation} are generally called strings and rotators \cite{Haase2021resourceefficient}, and they preserve the canonical commutation relations \eqref{standard_CR}, indeed
	\begin{equation}
		[R_{{\bf n}},P_{{\bf m}}]=\delta_{\bf{nm}}P_{{\bf n}}.
		\label{dual_CT}
	\end{equation}
	This reformulation acts only on the gauge degrees of freedom, therefore modifying the kinetic term of the matter Hamiltonian and not affecting the mass term. After the reformulation, Gauss laws are imposed to reduce the number of degrees of freedom by removing open strings while retaining the closed loops as the physical variables associated to the gauge fields \cite{Rothe,Creutz1977,Creutz1985}. As shown in Fig. \ref{2d_lattice_vars}, the physical states include the plaquette variables, $P_{{\bf n}}$, and the large horizontal and vertical loops wrapping the torus, which we denote by $\mathcal{L}_{\hat{x},\hat{y}}$. 
	
	For the minimal torus, we have eight links and three independent Gauss laws, one for each lattice site but the reference site ${\bf n}_R=(0,1)$. Solving the Gauss laws amounts to rephrase the theory in terms of five rotators and strings, the latter being three plaquettes and two wrapping loops. Moreover, in the pure gauge case the loops $\mathcal{L}_\mu$ commute with the Hamiltonian, $[H_G,\mathcal{L}_\mu]=0$, and therefore are constants of motion. As they vanish in the ground state, we set $\mathcal{L}_{\hat{x},\hat{y}}=0$ \cite{Haase2021resourceefficient}. The magnetic and electric contributions to the dual gauge Hamiltonian become \cite{Haase2021resourceefficient}
	\begin{equation}
		H_B=\frac{1}{2g^2}\bigg(8-\sum_{{\bf n}\neq{\bf n}_R}P_{{\bf n}}-\prod_{{{\bf n}\neq{\bf n}_R}}P_{{\bf n}}+\text{H.c.}\bigg)
		\label{minimal_torus_HB},
	\end{equation}
	\begin{equation}
		H_E=2g^2\bigg(\sum_{{\bf n}\neq{\bf n}_R}R_{{\bf n}}^2-R_{(1,0)}[R_{(0,0)}+R_{(1,1)}]\bigg).
		\label{minimal_torus_HE}
	\end{equation}

    We emphasize that the dual Hamiltonian $H=H_B+H_E$ can be split into a local and a nonlocal part, defined as
	\begin{equation}
		H_{\text{loc}}=2g^2\sum_{{\bf n}\neq{\bf n}_R}R_{{\bf n}}^2+\frac{1}{2g^2}\sum_{{\bf n}\neq{\bf n}_R}[2-P_{{\bf n}}-P_{{\bf n}}^\dagger],
		\label{H_local}
	\end{equation}
	\begin{align}
		\nonumber
		H_{\text{nonloc}}=&-2g^2R_{(1,0)}(R_{(0,0)}+R_{(1,1)})\\
		&\frac{1}{2g^2}(2-P_{(0,0)}P_{(1,0)}P_{(1,1)}+\text{H.c.}).
		\label{H_nonLocal}
	\end{align}
    This separation allows for a systematic many-body treatment of LGTs and its importance will be further discussed in the next Section. We refer to the Appendix \ref{PBC_OBC_details} for the generalization to $N\times N$ lattices, both for periodic and open boundary conditions, and in the presence of dynamical matter. 

    \subsection{\label{LocalBasisU1} Identification of the optimal local basis}
    \subsubsection{\label{RDBU1} The single-plaquette Hamiltonian in $\text{U}(1)$ LGTs}
        
    The first step in the use of the RDB for a lattice computation is the resolution of the single-plaquette problem. We provide here the main points of the process, while referring to Appendix \ref{local_basis_computations} for computational details and a more mathematically refined exposition.  

    In the KS formulation, the single-plaquette problem consists of a gauge degree of freedom on each of the four links in a square. After dualization and gauge fixing, only one gauge degree of freedom remains \cite{Haase2021resourceefficient}. The Hilbert space of this remaining degree of freedom is the space of square-integrable functions over the group, $L^2(\text{U}(1))$, which is infinite-dimensional. To digitize the theory, we need to select a basis of $L^2(\text{U}(1))$ in order to truncate to a finite-dimensional subspace. One possible choice is the electric basis, consisting of eigenstates of the electric Hamiltonian (see Appendix \ref{local_basis_computations} for more details), that works at strong coupling but fails when approaching the weak coupling limit. Instead, we use the eigenstates of the single-plaquette Hamiltonian to obtain a basis of $L^2(\text{U}(1))$ adapting to the competition between electric and magnetic fields. The single-plaquette Hamiltonian for a pure $\text{U}(1)$ LGT is
    \begin{equation}
        H_{\text{SP}}=2g^2R_{{\bf}}^2+\frac{1}{2g^2}[2-P_{{\bf}}-P_{{\bf}}^\dagger],
        \label{single_plaquette_hamiltonian}
    \end{equation}
    in terms of the plaquette operator $P$ and its conjugate electric field $R$, as described in the previous Section. 
    We parametrize the plaquette $P=\exp(i\theta)$ in terms of the group coordinate $\theta\in[0,2\pi)$ and expand a generic state in $L^2(\text{U}(1))$ in terms of group elements, i.e.,
	\begin{equation}
		|\Psi\rangle = \int_0^{2\pi}\frac{d\theta}{2\pi}\Psi(\theta)|e^{i\theta}\rangle,
		\label{state_group_basis}
	\end{equation}
	where we used the Haar measure of the group $d\mu(\theta)=(2\pi)^{-1}d\theta$.
        
    Due to the commutation relations in Eq. \eqref{dual_CT}, the rotator turns into a differential operator $R=-i\partial_{\theta}$. In this coordinate representation, the single-plaquette Hamiltonian is given by the differential operator
	\begin{equation}
		H_{\text{SP}}(g)=-2g^2\partial^2_{\theta}+\frac{1-\cos\theta}{g^2},
		\label{H0_group_coordinate_reps}
	\end{equation}
	and the single-plaquette problem is defined by the Schroedinger equation $H_{\text{SP}}(g)|\Psi\rangle=E|\Psi\rangle$, that can be written as
	\begin{equation}
		\bigg[-2g^2\partial_{\theta}^2+\frac{1}{g^2}(1-\cos\theta)-E\bigg]\Psi(\theta) = 0.
		\label{H0_diffeq}
	\end{equation}
	The above equation is of the Mathieu form \cite{Ligterink2000}, admitting solutions in terms of Mathieu functions, whose parameters are related to the bare coupling and energies of the theory. We refer to Appendix \ref{local_basis_properties} for the full characterization of its properties, and the analytical discussion of the limiting cases of weak and strong coupling regimes. 

    The single-plaquette problem of Eq. \eqref{H0_diffeq} with fixed coupling strength $g$ provides a basis of states $|\Psi_\alpha\rangle$ of $L^2(\text{U}(1))$ ordered in increasing energies by $\alpha\in\mathbb{N}$. By solving Eq. \eqref{H0_diffeq} for different coupling strengths we obtain a $g-$dependent basis $|\Psi_\alpha(g)\rangle$ interpolating between weak and strong coupling. In practice, only the lowest-lying $L_{\text{max}}+1$ states are computed. The knowledge of the operators $R$ and $P$ is needed when addressing the full lattice problem. After computing the first low-lying states $|\Psi_\alpha(g)\rangle$ we compute the matrix elements for all such operators in the truncated subspace. Therefore, all computations needed to write the full lattice Hamiltonian in the tensor-product of the single-plaquette eigenbasis are only performed once, before working on the targeted lattice problem.
	
	\subsubsection{\label{app_RDB}Variational basis optimization for $\text{U}(1)$ LGTs}
    Once the single-plaquette Hamiltonian and all necessary matrix elements have been precomputed, the basis can be used for any lattice computation. We exemplify the procedure for the minimal torus lattice described in Section \ref{DualLBO}. As discussed previously, in the ground state of such a system the relevant degrees of freedom are the 3 independent plaquettes $P_{{\bf n}}$. We control the basis selected for each plaquette by the couplings $g_{\bf n}$, i.e., the basis $|\Psi_\alpha(g_{\bf n})\rangle$ is used to span the Hilbert space of the $\bf n$ plaquette. To distinguish the parameters $g_{\bf n}$ that label the basis from the coupling $g$ in the Hamiltonian we indicate them in the following as basis parameters. To ease the notation we indicate the local parameter vector as $\bf{g}$, i.e., ${\bf{g}} = (g_{(0,0)},g_{(1,0)},g_{(1,1)})$ for the minimal torus.

    We perform the truncation in the local basis by retaining $L_{\text{max}}+1$ states, i.e., by selecting all the wave functions with $\alpha_{\bf n}\leq L_{\text{max}}$ (we start counting the states from 0, which labels the ground state). The full system is then truncated to the subspace generated by the tensor product of the local bases and the Hamiltonian is projected onto this subspace. An essential step in the RDB is the optimal selection of the local bases for each coupling $g$ and truncation $L_{\text{max}}$. By reabsorbing the change of basis to the (matrix elements of the) Hamiltonian, i.e., $H\rightarrow H_{g,L_{\text{max}}}(\bf{g})$, the RDB can equivalently be seen as a method to obtain the optimal Hamiltonian representation at each $g$ and $L_{\text{max}}$. 
    
    The local Hamiltonian in Eq. \eqref{H_local} is the sum of three independent single-plaquettes terms, $H_{\text{loc}}=\sum_{{\bf n}\neq{\bf n}_R} H_{\text{SP},\bf{n}}$. Therefore, reasonable values for the basis parameters can be taken to be ${\bf g}=g\;(1,1,1)$. In the following we refer to this choice as dual basis. However, such a procedure would ignore the effects of nonlocal terms in the determination of the basis parameters. 
    
    To find the optimal local basis at fixed truncation $L_{\text{max}}$ and coupling $g$ we proceed in two steps. First, we diagonalize $H_{g,L_{\text{max}}}(\bf{g})$ and obtain its ground state energy $E_{g,L_{\text{max}}}(\bf{g})$. Second, we interpret $E_{g,L_{\text{max}}}(\bf{g})$ as a cost function to be minimized. Thus, the optimal basis is identified by $\bf{g}_{\text{RDB}}$ satisfying 
    \begin{equation}
        E_{g,L_{\text{max}}}({\bf g}_{\text{RDB}})=\min_{\bf g}E_{g,L_{\text{max}}}({\bf g}).
    \end{equation}
    Equivalently, $H_{g,L_{\text{max}}}(\bf{g}_{\text{RDB}})$ represents an optimal Hamiltonian representation for coupling $g$ and number of resources given by $L_{\text{max}}$.
	
	\subsubsection{\label{symmetries_truncation}Symmetries}
	The single-plaquette Hamiltonian \eqref{H0_group_coordinate_reps} has a $\mathbb{Z}_2$ parity symmetry under the transformation $\theta_{\bf n}\rightarrow-\theta_{\bf n}$, resulting in $P_{\bf n}=\exp{(i\theta_{\bf n})}\rightarrow\exp{(-i\theta_{\bf n})}=P^\dagger_{\bf n}$ and $R_{\bf n}=-i\partial_{\theta_{\bf n}}\rightarrow i\partial_{\theta_{\bf n}}=- R_{\bf n}$. Therefore, the single-plaquette basis states can always be taken to be even or odd functions. While this symmetry is not preserved at the level of individual plaquettes in the many-body Hamiltonian, the latter is invariant under the joint parity transformation of the three loops. This can be seen by inspecting directly the nonlocal terms in Eq. \eqref{H_nonLocal}. Enforcing this global symmetry approximately halves the dimension of the Hilbert space. 	
	To further optimize the computational basis, we impose an additional truncation and limit the total number of possible single-plaquette excitations by setting $\alpha_{\bf n_1}+\alpha_{\bf n_2}+\alpha_{\bf n_3} \leq N_{\text{max}}$. Combined with the parity symmetry, this procedure provides a systematic truncation scheme that depends on both $L_{\text{max}}$ and $N_{\text{max}}$. The improvement achieved with this further optimization can be clearly seen in Fig. \ref{relative_plaq_vs_Hilbert_space_size}. More results on the performance of the global truncation are shown in Appendix \ref{global_truncation}.

	\section{\label{basis_performance}Performance of the renormalized dual basis}
	In this Section we evaluate the performance of the proposed computational basis in terms of resource efficiency, estimating the retained number of states for the computation of physically relevant observables. We compare our protocol with existing ones developed for the study of $\text{U}(1)$ LGTs and cQED in 2+1D, showing that our computational basis significantly reduces the required resources needed to attain a given level of precision.
    
	\subsection{\label{hilbert space}Hilbert space size}
	As a first benchmark, we evaluate the different bases in the computation of the expectation value of the plaquette operator, defined as
	\begin{equation}
		\langle\square\rangle\equiv\frac{g^2}{2N_{\text{plaq}}}\langle\psi_0|H_B|\psi_0\rangle,
		\label{HB_expval_definition}
	\end{equation}
    where $N_{\text{plaq}}$ is the number of plaquettes in the system.
    
	This quantity is of fundamental importance, as it can be related to the running of the coupling, and its continuum extrapolation stands as an open problem in the study of LGTs \cite{CreutzSU(2)1980,Haase2021resourceefficient,PaulsonPRX2021,ClementePRD2022}.
	
	\begin{figure}[t!]
		\centering
		\includegraphics[width=1.0\linewidth]{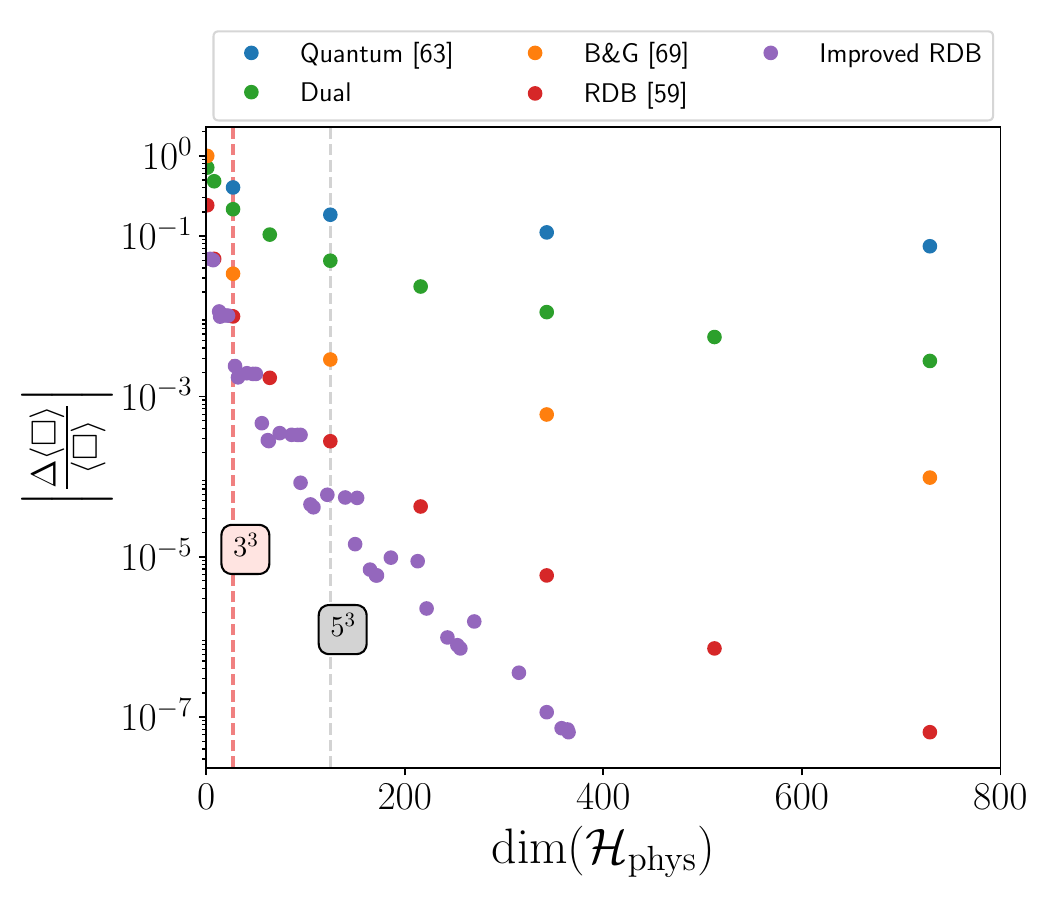}
		\caption{Relative precision on the plaquette expectation value for the 2+1D $\text{U}(1)$ LGT on the minimal torus as a function of the total size of the Hilbert space. We compare different formulations and basis choices: “Quantum” (blue circles) \cite{Haase2021resourceefficient}, “Dual" (green circles), “B\&G” (orange circles) \cite{BauerPRD2023}, “RDB” (red circles) \cite{Fontana2024}, and “Improved RDB” (purple circles), see Sec. \ref{hilbert space}. The dashed vertical light coral and grey lines highlight the physically relevant cases of $\text{dim}(\mathcal{H}_{\text{phys}})=27,\;125$, corresponding to three and five retained states per plaquette, respectively.}
		\label{relative_plaq_vs_Hilbert_space_size}
	\end{figure}

    \begin{figure*}[t!]
		\centering
		\includegraphics[width=1.0\linewidth]{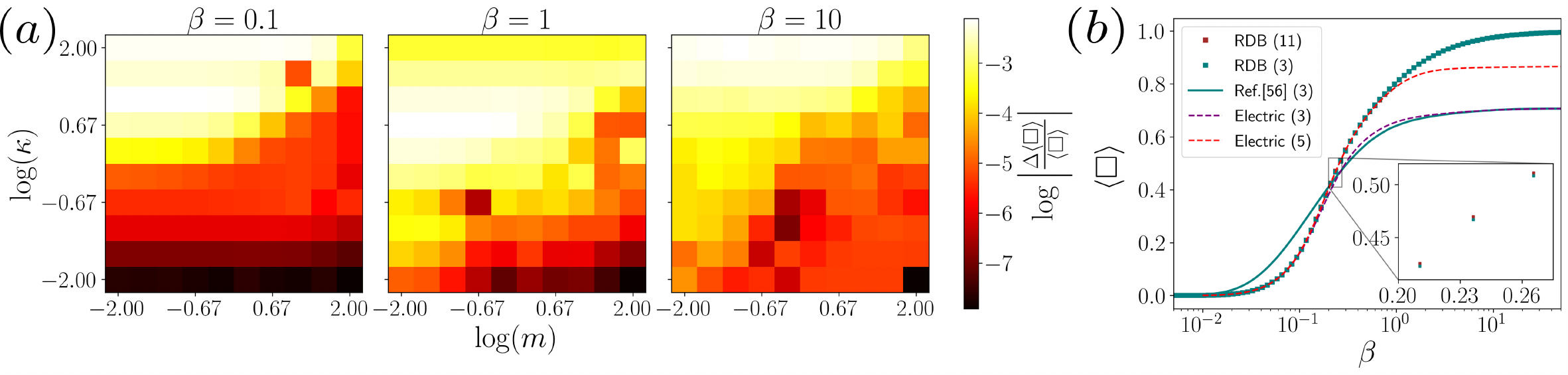}
		\caption{\textbf{Application of the RDB to cQED}. (a) Logarithm of the relative precision on the plaquette expectation value for the single plaquette cQED, obtained with the RDB truncated to $L_{\text{max}}=2$ (3 gauge states), as a function of the fermion mass $m$ and kinetic energy $\kappa$. The exact result has been taken to be the highest truncation computed, $L_{\text{max}}=10$ (11 gauge states). Three different coupling strengths are used: strong coupling ($\beta=0.1$), intermediate coupling ($\beta=1$) and weak coupling ($\beta=10$). (b) Plaquette expectation value for the 2+1D cQED on a single plaquette as a function of $\beta=\frac{1}{2g^2}$. We compare the formulation of \cite{Haase2021resourceefficient} used in a recent trapped-ion computation \cite{Meth2023Oct} (from where we extract the data), with the RDB for the truncations $L_{\text{max}}\in\{2,10\}$. The results with $L_{\text{max}}=10$ is taken as exact and the inset shows the biggest error when the gauge field is truncated to a qutrit. Results obtained by truncating to the electric basis have been added to assist in the comparison. The gauge sector Hilbert space dimension is indicated in parentheses.}
		\label{QEDcomparison}
	\end{figure*}
	We compute the relative precision $|\Delta\langle\square\rangle/\langle\square\rangle|$ as a function of the total Hilbert space dimension $\text{dim}(\mathcal{H}_{\text{phys}})$ in five different cases: the two protocols presented in Refs. \cite{Haase2021resourceefficient} (Quantum) and \cite{BauerPRD2023} (B\&G) based on the discretization of $\text{U}(1)$ to $\mathbb{Z}_N$, the single-plaquette Hamiltonian basis with (RDB) or without (Dual) variational optimization, and the global truncation presented in Section \ref{symmetries_truncation} (Improved RDB). We show our results in Fig. \ref{relative_plaq_vs_Hilbert_space_size}, and the comparison shows clearly that the RDB-based protocols perform better, i.e., they require less basis states to achieve the same precision, with respect to other basis choices. As a concrete resource comparison for the minimal torus, we consider the cases of $\text{dim}(\mathcal{H}_{\text{phys}})=27,\;125$, i.e., three and five retained states per independent plaquette. These two particular cases correspond to the qutrit and ququint encodings for the plaquette fields, recently employed to realize the first trapped-ion quantum computation for the 2+1D $\text{U}(1)$ LGT on the minimal torus \cite{Meth2023Oct}. As highlighted in Fig. \ref{relative_plaq_vs_Hilbert_space_size}, in the experimentally considered cases we are able to improve by two orders of magnitude the relative precision on the plaquette expectation value with respect to the existing schemes. 
    
    We believe that this improvement is of fundamental importance towards the large-scale quantum simulation and computation of LGTs: the RDB drastically reduces the required number of states needed to achieve a given precision, thus allowing for accurate estimates of physical observables while optimizing the resources. The difference between the compared computational bases is indeed more evident for larger sizes of the physical Hilbert space. By fitting the scaling of $Y=\log\left|\Delta\langle\square\rangle/\langle\square\rangle \right|$ with respect to $X=\log{\text{dim}(\mathcal{H}_{\text{phys}})}$, we find that the “Quantum” scheme exhibits linear scaling, while both the “B\&G” and RDB-based approaches (normal and improved) follow a power-law behavior, $Y \sim -X^b$. Our methods show power-laws $Y\propto -X^{2.5}$ (RDB) and $Y\propto -X^{3.2}$ (Improved RDB), while “B\&G” shows power-law $Y\propto -X^{2.1}$. Further details on the fitting procedure and parameter estimates are provided in Appendix \ref{scaling_resources}.

	\subsection{\label{cQED_OBC_2x2}Single plaquette compact quantum electrodynamics}
    
    To demonstrate the convenience of the RDB, we turn to the minimal example of cQED by including fermions to a single plaquette with open boundary conditions. We apply the RDB to the efficient reformulation provided in \cite{Haase2021resourceefficient}. The phase diagram of the system is governed by the strength of the gauge field coupling $\beta$, the fermion hopping parameter $\kappa$ and the mass of the fermions $m$. Fig. \ref{QEDcomparison}(a) shows the relative error in the plaquette expectation value obtained with a gauge-field truncation $L_{\text{max}}=2$, evaluated across the parameter space. Reference values are computed using the RDB with $L_{\text{max}}=10$, which yields converged results and serves as the benchmark. Although these results are limited to a small lattice, the achieved accuracy provides preliminary evidence of the ability of the RDB to adapt to the presence of matter terms in the Hamiltonian.
    
    Notably, this setup was recently implemented in a trapped-ion architecture \cite{Meth2023Oct} following the resource-efficient formulation of \cite{Haase2021resourceefficient}. In that experiment, the authors measured the plaquette observable for the specific case $\kappa=5$ and $m = 0.1$. We show in Fig. \ref{QEDcomparison}(b) the curve that would be obtained, in the absence of experimental error, with the formulation used in the trapped-ion simulation. For comparison, we also plot the curves obtained with the RDB in two cases: for the largest truncation considered ($L_{\text{max}}=10$, taken to be the exact curve) and for $L_{\text{max}}=2$, turning the gauge field into a qutrit, as used in \cite{Meth2023Oct}.
    While further comparison will require the design of an experimental proposal, even at the level of the local Hilbert space dimension it is clear that the RDB provides significantly higher accuracy with the same computational resources.
    \begin{figure*}[t!]
		\centering
		\includegraphics[width=1.0\linewidth]{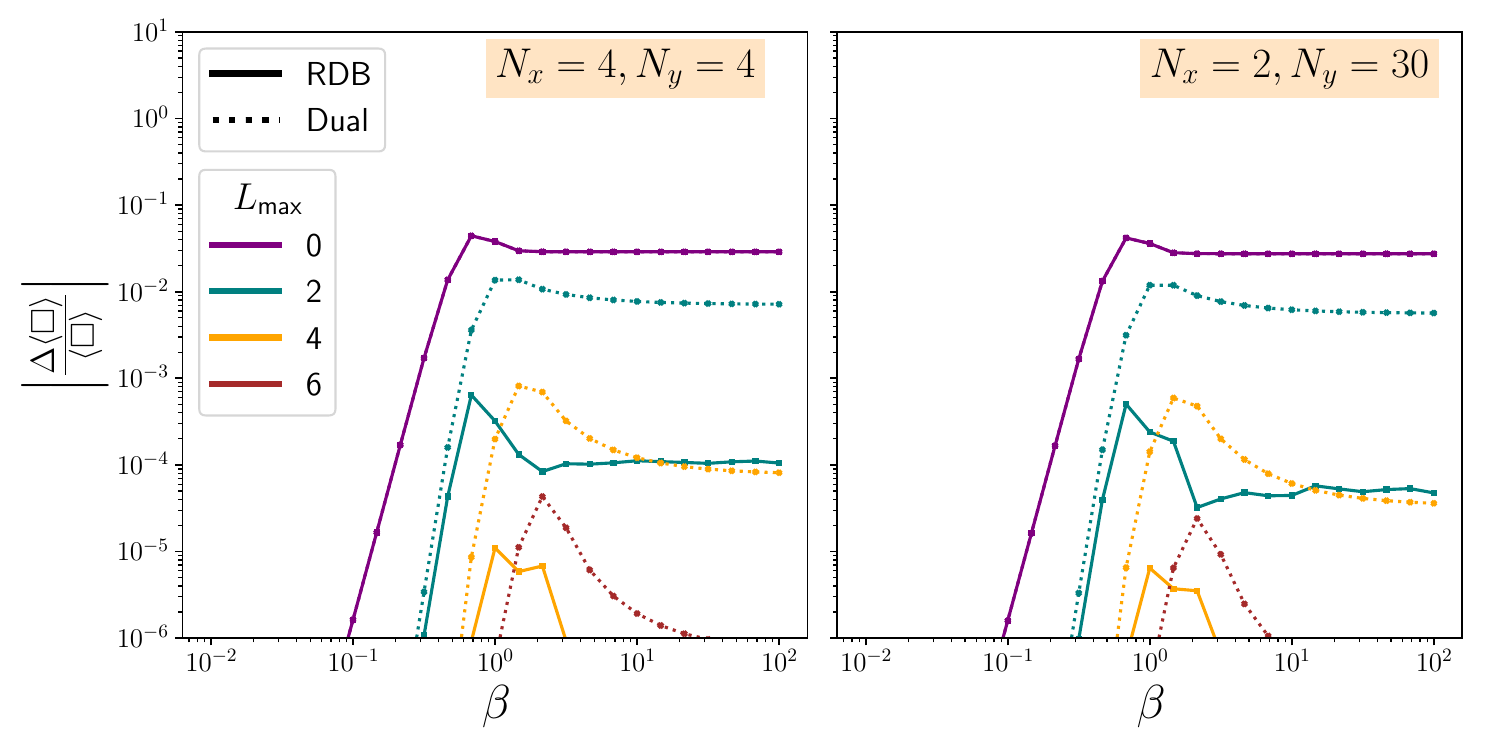}
		\caption{Relative precision on the plaquette expectation value for the 2+1D $\text{U}(1)$ LGT on a $4\times4$ lattice (left) and a $2\times30$ lattice (right), as a function of $\beta=\frac{1}{2g^2}$. In both cases the exact result are taken to be the one minimizing the energy at each $\beta$.  For the $4\times4$ lattice these were either $L_{\text{max}}=5$($8$) with (without) variational optimization. For the $2\times30$ lattice these are either $L_{\text{max}}=7$($9$) with (without) variational optimization.}
		\label{tensorNet}
    \end{figure*}
    
	\section{\label{large_scale_section}Large scale simulations for the pure gauge Hamiltonian}
    Tensor network algorithms have proven to be powerful tools for the study of LGTs \cite{RicoPRL2014,Silvi_2014,banulsRPP2020,Banuls2020,magnifico2024preprint}. In the Schwinger model, the matrix product state (MPS) ansatz has been employed to compute low-energy mass excitations \cite{Banuls2013-rb}, parton distribution functions \cite{bañuls2025partondistributionfunctionsschwinger}, and to investigate relevant phenomena such as confinement and string breaking \cite{PichlerPRX2016,osborne2023probingconfinement,BelyanskyPRL2024,PapaefstathiouPRD2025}. While the MPS ansatz is most efficient for 1+1D systems, it can also be applied to intermediate-sized 2+1D lattices by mapping them onto effective 1+1D problems. 
    
    In this Section, we use the MPS ansatz and the density-matrix renormalization group \cite{DMRG1992} to compute the ground state of a pure $\text{U}(1)$ LGT on larger lattices and demonstrate the effectiveness of the RDB in large-scale simulations \footnote{Tensor network computations have been performed with the TeNPy Library \cite{tenpy2024}. The bond dimension $\chi$ of the MPS ansatz was restricted to $\chi<4096$}. We have employed the dual Hamiltonian provided by \cite{Haase2021resourceefficient, Bender-Zohar} for a pure $\text{U}(1)$ LGT in open boundary conditions.  When variational minimization is applied, the same variational parameter is used for all plaquettes of the lattice. Although the use of different variational parameters would handle boundary effects more efficiently, the simpler choice of a shared variational parameter already yields fast convergence. In Fig. \ref{tensorNet} we illustrate the performance of the method in two representative settings, showing the relative precision of the plaquette operator for different truncations on lattices of sizes $4\times4$ and $2\times 30$. As the figure shows, the RDB remains efficient when scaling the system size, and up-scaling the lattice does not require an exponentially growing local Hilbert space. In particular, a truncation to only three local basis states is sufficient to achieve per-mille level relative accuracies for the plaquette operator.
    
	\section{\label{conclusions}Conclusions}
    In this work, we have demonstrated the superior performance of the RDB over existing methods in proof-of-principle calculations, as well as its scalability to larger lattices where, with previously available methods, neither exact diagonalization nor tensor network Hamiltonian approaches were able to operate reliably across all couplings. To our knowledge, the only previous 2+1D tensor network studies of a $\text{U}(1)$ LGT employ the quantum link model formalism, which is intrinsically limited in the weak-coupling regime \cite{Tschirsich19, Felser2020}.

    The success of the RDB is rooted in its definition. The dualization procedure converts basis optimization into a local problem and yields a dual Hamiltonian that naturally separates into local and nonlocal contributions. The single-plaquette Hamiltonian readily captures the local interactions, while the variational optimization effectively renormalizes the local basis due to the nonlocal interactions. Furthermore, since the single-plaquette Hamiltonian incorporates the magnetic interactions, the RDB is intrinsically designed to work for any value of the coupling. In contrast to electric-like truncation schemes, we avoid the exponential growth of the Hilbert space when approaching the physically relevant weak coupling limit. Finally, the one-time precomputation of the single-plaquette problem substantially reduces the computational cost for subsequent simulations. 

    The ultimate goal of the LGT community is largely to compute the phase diagram and dynamical processes in 3+1D quantum chromodynamics. By overcoming the roadblock of scalability, this work advances in the journey towards this goal and highlights  RDB as the path for both quantum computers and quantum-inspired algorithms. 

    Our results show that the RDB effectively reduces the complexity of simulating $\text{U}(1)$ theories -- across all coupling regimes -- to that of $\mathbb{Z}_N$ models, immediately enabling precise numerical investigations of key physical phenomena such as confinement, dynamical time evolution, and the roughening transition in 2+1D with quantum processors like in \cite{Meth2023Oct, rosanowski202521dquantumelectrodynamicsfinite}, and with tensor networks, as already demonstrated for $\mathbb{Z}_N$ \cite{Robaina2021,Wu2025, dimarcantonio2025rougheningdynamicselectricflux}. 

    In fact, by combining the RDB with algorithms tailored for two-dimensional systems, e.g., infinite projected entangled pair states \cite{verstraete2004renormalizationalgorithmsquantummanybody,Jordan2008,Phien15,Naumann2024}, we can achieve the continuum limit with ab initio tensor network calculations \footnote{Marc Miranda-Riaza, Pierpaolo Fontana, Alessio Celi (in preparation)}. While the same computation with dynamical charges and  non-Abelian gauge groups present additional challenges \cite{Fontana2024}, we are currently working to circumvent them by leveraging the modularity of the RDB approach.

    Before approaching the continuum limit, the application of the RDB to study SU(3) LGTs on small and intermediate-scale lattices is certainly an essential, and currently affordable step towards quantum chromodynamics we are currently pursuing. Beyond the higher complexity of the gauge sector, a major challenge of quantum chromodynamics is color and flavour degeneracy. Improving the RBD with Hamiltonian truncation methods established in the continuum \cite{HTYurov1990,HTHogervorst2015, HTHoutz2022, HTMiro2023} and recently applied to gauge theories \cite{HTHoutz2025, HTFitzpatrick2025} can mild this challenge.

    Last but not least, the RDB has the potential to boost real-time dynamics in quantum processors, beyond one spatial dimensions and purely electric dynamics \cite{Meth2023Oct,GonzalezCuadra2025,crippa2024analysisconfinementstring2}. In this context, combining the RDB with information-theoretic resource theory could help identify the regimes in which a genuine quantum advantage may emerge in LGT simulations \cite{Tarabunga23, Falcao25, santra2025quantumresourcesnonabelianlattice, ebner2025magicbarrierthermalization}. Ultimately, the RDB opens the door to high-precision quantum simulation of gauge theories beyond one spatial dimension, with the potential to discover new physical phenomena \cite{Bernien2017,mark2025observationballisticplasmamemory} also in higher dimensions.    

	\section{Acknowledgements}
	We acknowledge funding from MCIN/AEI/10.13039/501100011033 (MAPS PID2023-149988NBC21) and Generalitat de Catalunya (AGAUR 2021 SGR 00138). M. M. acknowledges support from MICINN/10.13039/501100004837 through an FPU grant (FPU22/03907). P.F. acknowledges support of the program Investigo (ref. 200076ID6/BDNS 664047), funded by the European Union through the Recovery, Transformation and Resilience Plan NextGenerationEU. All authors acknowledge funding from EU QUANTERA DYNAMITE (funded by MICN/AEI/10.13039/501100011033 and by the European Union NextGenerationEU/PRTR PCI2022-132919 (Grant No. 101017733)) and from the Ministry of Economic Affairs and Digital Transformation of the Spanish Government through the QUANTUM ENIA project call Quantum Spain project.
	
	\appendix
	\widetext
	
	\section{\label{PBC_OBC_details}Encoded Hamiltonian for arbitrary lattices and boundary conditions}
	We generalize the expressions in Eqs. \eqref{minimal_torus_HB}, \eqref{minimal_torus_HE} to arbitrary $N\times N$ lattices, discussing as well the inclusion of dynamical matter. We discuss first the difference between open and periodic boundary conditions, summarized in Table \ref{table1}. In the KS formulation, there are $2$ links for each lattice site, for a total of $2N^2$ Wilson operators. Due to the lattice topology, there are $N^2-1$ independent Gauss laws, one for each site apart the reference site ${\bf n}_R$ (see Fig. \ref{2d_lattice_vars}). Therefore, the number of physical degrees of freedom is $N^2+1$, equal to the number of dual degrees of freedom needed to describe the theory, namely $N^2-1$ plaquettes and $2$ large loops. In the open case, we have $2N^2-2N$ lattice links and $(N+1)^2-1$ Gauss laws, resulting in $N^2$ physical degrees of freedom. This is exactly the number of total lattice plaquettes in the dual formulation, without residual constraints left \cite{Bender-Zohar,Haase2021resourceefficient,Fontana2022:LGT}.
	
	\subsection*{Periodic boundary conditions}
	Before writing the full expression of the dual Hamiltonian in presence of dynamical matter, we briefly comment about two technical points. First, as a consequence of the zero total flux condition, due to periodic boundary conditions, the plaquette $P_{(0,N-1)}\equiv P_{{\bf n}_R}$ can be expressed as the product of all the others
	\begin{equation}
		P_{{\bf n}_R}^\dagger=\prod_{{\bf n}\neq {\bf n}_R}P_{{\bf n}}
		\label{P_reference_site}
	\end{equation}
	generating an interacting term in the dual magnetic Hamiltonian \cite{StrykerPRD2020}. Correspondingly, it is convenient to set $R_{{\bf n}_R}=0$. Second, given the arbitrariness in the resolution of the Gauss law, we need to fix a prescription for its lattice implementation. Following the convention of \cite{Haase2021resourceefficient}, depicted in the right lattice in Fig. \ref{2d_lattice_vars}, we connect each charge operator $q_{{\bf n}}$ to the origin with the string $(0,0)\rightarrow(n_x,0)\rightarrow(n_x,n_y)$, resulting in the corrections $q_{{\bf n},\mu}$ to the electric field $E_{\mu}({\bf n})$ defined by
	\begin{equation}
		q_{{\bf n},\hat{x}}\equiv-\sum_{m_x=n_x+1}^{N-1}\sum_{m_y=0}^{N-1}\delta_{n_y,0}q_{{\bf m}},
		\label{q_x_def}
	\end{equation}
	\begin{equation}
		q_{{\bf n},\hat{y}}\equiv-\sum_{m_x=0}^{N-1}\sum_{m_y=n_y+1}^{N-1}\delta_{n_x,m_x}q_{{\bf m}}.
		\label{q_y_def}
	\end{equation}
	
	Given these considerations, the dual magnetic and electric Hamiltonians become \cite{Haase2021resourceefficient}
	\begin{equation}
		H_B=-\frac{1}{2g^2}\bigg[N^2-\sum_{{\bf n}\neq{\bf n}_R}P_{{\bf n}}-\prod_{{\bf n}\neq {\bf n}_R}P_{{\bf n}}\bigg]+\text{H.c.},
		\label{PBC_square_lattice_HB}
	\end{equation}
	\begin{equation}
		H_E=\frac{g^2}{2}\sum_{{\bf n}}\bigg[(\delta_{n_y,0}R_{\hat{x}}+R_{{\bf n}}-R_{{\bf n}-\hat{y}}+q_{{\bf n},\hat{x}})^2+(\delta_{n_x,0}R_{\hat{y}}+R_{{\bf n}-\hat{x}}-R_{{\bf n}}+q_{{\bf n},\hat{y}})^2\bigg],
		\label{PBC_square_lattice_HE}
	\end{equation}
	while the kinetic term is
	\begin{equation}
		H_k=\kappa\sum_{{\bf n}}\psi^\dagger_{{\bf n}}\bigg[\mathcal{L}_{\hat{x}}^{-\delta_{n_x,N-1}}\prod_{m_y=0}^{n_y-1}P_{(n_x,m_y)}\psi_{{\bf n}+\hat{x}}\bigg(\mathcal{L}_{\hat{y}}\prod_{m_x=n_x}^{N-1}\prod_{m_y=0}^{N-1}P_{{\bf m}}\bigg)^{\delta_{n_y,N-1}}\psi_{{\bf n}+\hat{y}}\bigg]+\text{H.c.}
		\label{PBC_square_lattice_Hk}
	\end{equation}
    
    \begin{table}
		\centering
		\setlength{\tabcolsep}{12pt}
		\renewcommand{\arraystretch}{1.3}
		\begin{tabular}{c|ccc}
			Boundary conditions & Links & Gauss laws & Physical DOFs  \\
			\hline
			Periodic & $2N^2$ & $N^2-1$ & $N^2+1$ \\
			Open & $2N(N+1)$ & $(N+1)^2-1$ & $N^2$ \\
		\end{tabular}
		\label{table1}
		\caption{Comparison of the counting of gauge degrees of freedom in the cases of periodic and open boundary conditions. We report the count of link variables in the Kogut--Susskind formulation, alongside the numbers of independent Gauss laws and physical degrees of freedom, which is obtained by resolving the local symmetry.}	
	\end{table}
	
	\subsection*{Open boundary conditions}
	The Hamiltonian in case of open boundary conditions is essentially the same of the previous Subsection, setting to zero the terms related to the global loops and rotators \cite{Bender-Zohar,PaulsonPRX2021}. We report its expression in the two cases that we treated in the main text: the tensor network simulations of square lattices, and the single plaquette cQED.
	
	\subsubsection{Pure gauge theory on $N\times N$ lattices} 
	We set $m=\kappa=0$, as well as all the contributions coming from the global loop and rotators, to obtain the pure gauge Hamiltonian
	\begin{equation}
		H_G=\frac{g^2}{2}\sum_{{\bf n}}\bigg[(R_{{\bf n}}-R_{{\bf n}-\hat{y}})^2+(R_{{\bf n}-\hat{x}}-R_{{\bf n}})^2\bigg]-\frac{1}{2g^2}\sum_{{\bf n}}(2-P_{{\bf n}}-P^\dagger_{{\bf n}}).
		\label{OBC_square_lattice_HG}
	\end{equation}
	\subsubsection{Compact quantum electrodynamics on a $2\times 2$ lattice}
	Following the convention of \cite{PaulsonPRX2021}, we can encode the gauge degrees of freedom in a single link, chosen to be the one joining $(0,1)\rightarrow(1,1)$. This results in the identification $P_{(0,0)}=U_{\hat{x}}(0,1)$ and its electric field $E_{\hat{x}}(0,1)$ as independent variables. Expressed in terms of them, the single-plaquette Hamiltonian with dynamical matter becomes
	\begin{equation}
		H_G=\frac{g^2}{2}\bigg[E^2_{\hat{x}}(0,1)+(E_{\hat{x}}(0,1)-q_{(0,1)})^2+(E_{\hat{x}}(0,1)-q_{(1,1)})^2+(E_{\hat{x}}(0,1)+q_{(0,0)}+q_{(0,1)})^2\bigg]-\frac{1}{2g^2}(U_{\hat{x}}(0,1)+U^\dagger_{\hat{x}}(0,1))
	\end{equation}
	\begin{equation}
		H_M=m\sum_{{\bf n}}(-1)^{n_x+n_y}\psi^\dagger_{{\bf n}}\psi_{{\bf n}}+\kappa\bigg(\psi^\dagger_{(0,0)}(\psi_{(0,1)}+\psi_{(1,0)})+\psi^\dagger_{(1,0)}\psi_{(1,1)}+\psi^\dagger_{(1,1)}U_{\hat{x}}(0,1)\psi_{(0,1)}+\text{H.c.}\bigg)
		\label{OBC_square_lattice_HM}
	\end{equation}
	
	\section{\label{local_basis_properties}Properties of the single-plaquette Hamiltonian}
	By changing variable to $\phi=\theta/2$, the Eq. \eqref{H0_diffeq} is in the form of Mathieu's differential equation
	\begin{equation}
		\bigg[\frac{d^2}{d\phi^2}+a-2q\cos(2\phi)\bigg]y(\phi)=0,
	\end{equation}
	with
	\begin{equation}
		a\equiv\frac{2}{g^2}\bigg(E^2-\frac{1}{g^2}\bigg),\qquad q\equiv-\frac{1}{g^4},.
		\label{mapping_mathieu_params}
	\end{equation}
	The solutions to this equation are called Mathieu functions, and are classified according to the parameters $a,\;q$ and the periodicity of $\phi$. In our case, we have $a,\;q\in\mathbb{R}$, and we can investigate everything as a function of $\phi$, by taking into account the corresponding change in the Haar measure, since $\phi\in[0,\pi)$ and $d\mu(\phi)=d\phi/\pi$.
	
	Considering the symmetries of the $\text{U}(1)$ group, we look for solutions that are periodic in $\theta$ on the circle, i.e., satisfying $\psi(\theta)=\psi(\theta+2\pi)$. The corresponding Mathieu functions must reflect this symmetry, and, since $2\phi\equiv\theta$ and the period $T_\theta=2\pi$, we have that $T_\phi=\pi$. Therefore, we have to select the Mathieu functions with period $\pi$ in the coordinate $\phi$. According to the tabulated values, the Mathieu functions of the first kind with period $T=\pi$ are
	\begin{equation}
		\text{ce}_n,\qquad n\;\text{even (even parity)},\qquad \text{se}_n,\qquad n\;\text{even (odd parity)}
		\label{mathieu_functions}
	\end{equation}
	For the ground state, where $n=0$, therefore we have
	\begin{equation}
		\psi_0(\theta)\propto \text{ce}_0\bigg(q=-\frac{1}{g^4},\phi=\frac{\theta}{2}\bigg)
	\end{equation}
	up to a normalization constant, at fixed values of the bare coupling $g$.
	
	\begin{figure}[t!]
		\centering
		\includegraphics[width=1.0\linewidth]{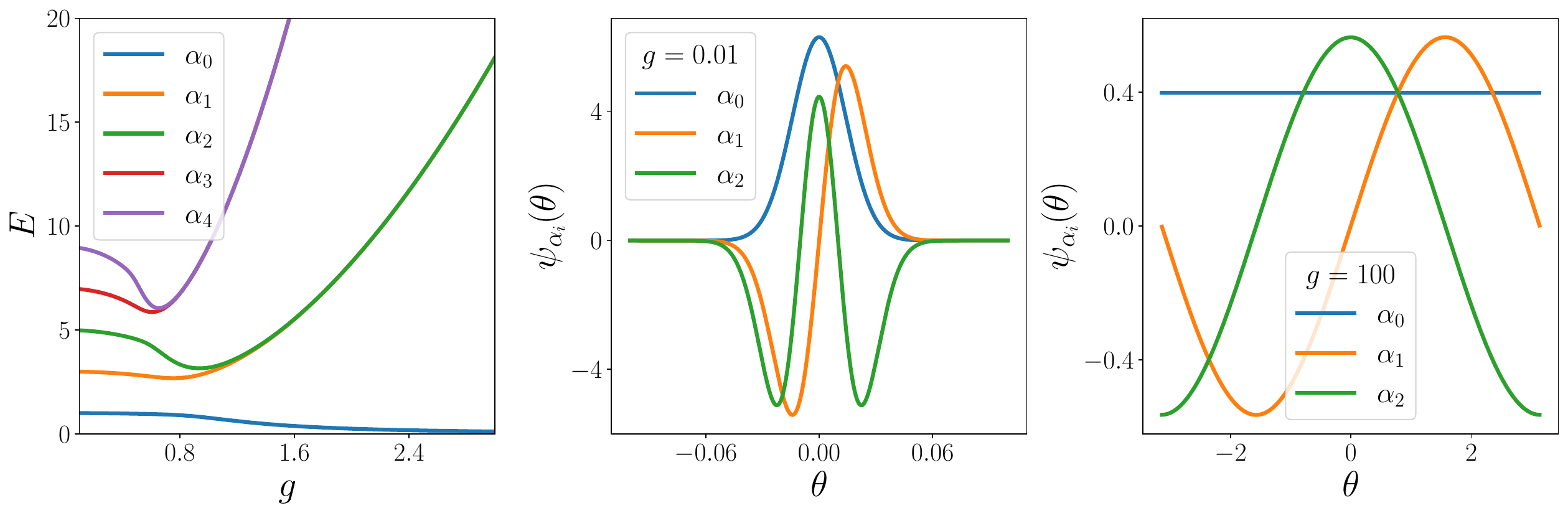}
		\caption{\textbf{Single-plaquette problem.} \textbf{Left}: lowest eigenvalues for the single-plaquette Hamiltonian as a function of the bare coupling $g$. \textbf{Center}: three lowest eigenstates at weak coupling, for $g=10^{-2}$. \textbf{Right}: three lowest eigenstates at strong coupling, for $g=10^2$. In the former, they are well approximated by the eigenstates of the harmonic oscillator. In the latter, they are well approximated by Fourier modes.}
		\label{spectrum_states_strong_weak}
	\end{figure}
	
	\subsection{\label{weak_coupling_Mathieu}Weak coupling limit}

	In this limit, $g\rightarrow 0$ and the potential term dominates in Eq. \eqref{H0_diffeq}. The support of the wave function is limited around $-g\lessapprox\theta\lessapprox g$, and we can expand the potential term around $\theta\simeq 0$. We have
	\begin{equation}
		1-\cos\theta=-\frac{\theta^2}{2}+O(\theta^4)\qquad\Rightarrow\qquad H_0(\theta)=\partial_\theta^2-\frac{\theta^2}{4g^2},
		\label{H0_weak_coupling}
	\end{equation}
	which is an harmonic oscillator of frequency $\omega=(\sqrt{2}g^2)^{-1}$. The energy levels are
	\begin{equation}
		E_n(\omega)=E_n(g^{-2})=\frac{1}{\sqrt{2}g^2}\bigg(n+\frac{1}{2}\bigg),
		\label{eigvals_weak_coupling}
	\end{equation}
	and wave functions can be expressed using Hermite polynomials
	\begin{equation}
		\psi_n(\theta)=\frac{1}{\sqrt{2^nn!}}\bigg(\frac{1}{\sqrt{2}g^2\pi}\bigg)^{\frac{1}{4}}e^{-\frac{\theta^2}{2\sqrt{2}g^2}}H_n\bigg(\frac{\theta}{2^{1/4}g}\bigg).
		\label{eigvecs_weak_coupling}
	\end{equation}
	For the ground state, we have simply $E_0=(2\sqrt{2}g^2)^{-1}$ and a gaussian profile
	\begin{equation}
		\psi_0(\theta)=\bigg(\frac{1}{\sqrt{2}g^2\pi}\bigg)^{\frac{1}{4}}e^{-\frac{\theta^2}{2\sqrt{2}g^2}},
		\label{GS_weak_coupling}
	\end{equation}
	as showed in the center plot of Fig. \ref{spectrum_states_strong_weak}.
	
	\subsection{\label{strong_coupling_Mathieu}Strong coupling limit}
	In this limit, the kinetic term dominates over the potential, and Eq. \eqref{H0_diffeq} reduces to
	\begin{equation}
		-2g^2\partial^2_\theta\psi(\theta)=E\psi(\theta)\qquad \Rightarrow\qquad -\partial_\theta^2\psi(\theta)=\tilde{E}\psi(\theta),\quad\tilde{E}\equiv\frac{E}{2g^2}.
	\end{equation}
	This is a free Schroedinger equation with periodic boundary conditions $\psi(\theta)=\psi(\theta+2\pi)$, a problem that is also known as the particle in a 1D ring. The energies $\tilde{E}$ are quantized because of the periodicity, and have values $\tilde{E}=n^2$, $n\in\mathbb{Z}$ (in this units, the mass of the particle is $m=1/2$). The wave functions are plane waves of the form
	\begin{equation}
		\psi(\theta)=\frac{e^{\pm in\theta}}{\sqrt{2\pi}}.
		\label{eigvecs_strong_coupling}
	\end{equation}
	Therefore, there are two degenerate quantum states for every $n\in\mathbb{N}^+$, corresponding to the two waves $e^{\pm in\theta}$. An equivalent approach would be to adopt the parity-preserving basis $\{\sin(n\theta),\cos(n\theta)\}_{n\in\mathbb{N}}$, which leads to the usual Fourier series and naturally connects with the strong coupling expansion. The ground state structure in this case is even simpler, since it correspond to the level $k_\theta=0$, i.e., $n=0$, with constant wave function in $\theta-$space (see the right plot of Fig. \ref{spectrum_states_strong_weak}).
	
	\section{\label{scaling_resources}Scaling behavior of the plaquette operator relative precision}
	\begin{figure}[t!]
		\centering
		\includegraphics[width=0.5\linewidth]{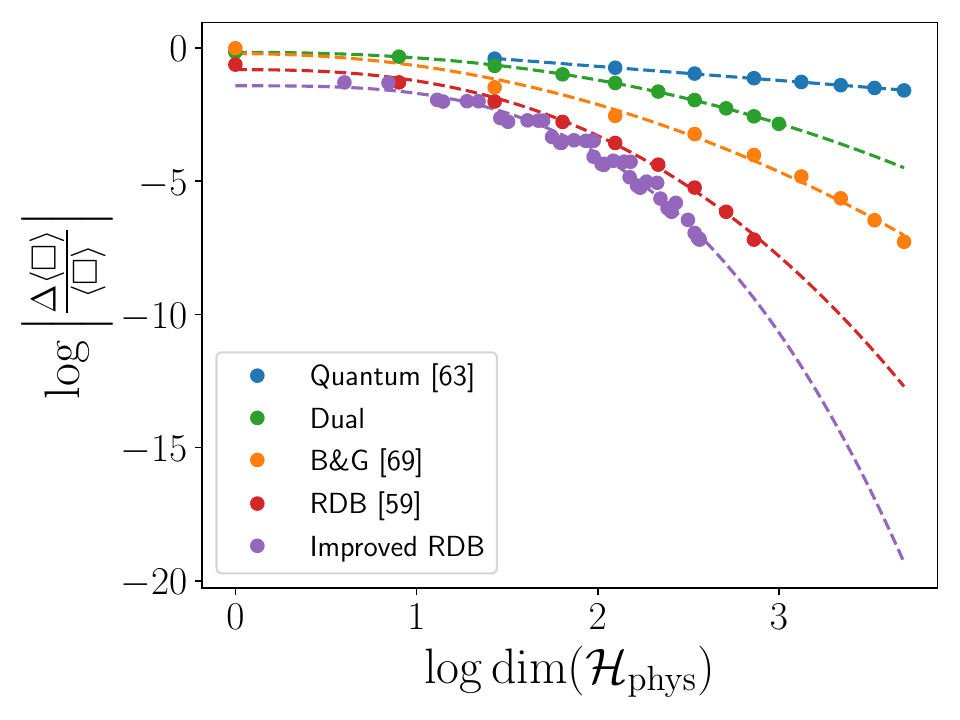}
		\caption{Logarithms of the relative precision of the plaquette expectation values and the total size of the physical Hilbert space. As in the main text, we compare the computational bases labeled as “Quantum” (blue circles) \cite{Haase2021resourceefficient}, “Dual” (green circles), “B\&G” (orange circles) \cite{BauerPRD2023}, “RDB” (red circles) \cite{Fontana2024} and “Improved RDB” (purple circles). We superimpose to the data sets the fit functions (straight line for the “Quantum” and Eq. \eqref{log_fit_function} for all the other sets) with dashed lines following the same color codes of the data.}
		\label{fig_scaling_relative_plaq_precision}
	\end{figure}
	We evaluate the behavior of the relative precision for the plaquette operator as a function of the physical Hilbert space size. To this purpose, it is better to visualize the data in Fig. \ref{relative_plaq_vs_Hilbert_space_size} in terms of their logarithms, as we plot in Fig. \ref{fig_scaling_relative_plaq_precision}. By defining,
	\begin{equation}
		X=\log{\text{dim}(\mathcal{H}_{\text{phys}})},\qquad Y=\log\left|\frac{\Delta\langle\square\rangle}{\langle\square\rangle}\right|
	\end{equation}
	we observe that all the data sets follow a power-law $Y\propto X^{\alpha}$, apart from the ``Quantum" data \cite{Haase2021resourceefficient}, clearly satisfying a linear relation $Y\propto X$. For the latter, we fit the data to the straight line $Y=aX+b$, obtaining the estimates $a=0.37(1)$, $b=-0.523(5)$. In terms of the original variables, this corresponds to the scaling law $\left|\Delta\langle\square\rangle/\langle\square\rangle\right|\approx [\text{dim}(\mathcal{H}_{\text{phys}})]^{1/3}$.
	
	Regarding the other data sets, we fit them with the function
	\begin{equation}
		Y=c-a X^b,\qquad b>0,
		\label{log_fit_function}
	\end{equation}
	obtaining the parameters reported in Table \ref{table2}. 
    For the RDB schemes we identify the power-laws $Y\propto X^{5/2}$ (RDB) and $Y\propto X^3$ (Improved RDB), within the parameters uncertainties. These numerical estimates show how the RDB allows improving the relative precision in the computation of the plaquette operator with respect to the existing schemes, of one order of magnitude (in log-scale) when including symmetry arguments in the basis implementation. We conclude therefore that, in terms of the original variables, the scaling with the Hilbert space dimension is
	\begin{equation}
		\left|\frac{\Delta\langle\square\rangle}{\langle\square\rangle}\right|\approx e^{-a\log^3{\text{dim}(\mathcal{H}_{\text{phys}})}}.
		\label{eq_scaling_relation_relative_plaq_vs_dim_Hp}
	\end{equation}
	\begin{table}
		\centering
		\setlength{\tabcolsep}{12pt}
		\renewcommand{\arraystretch}{1.3}
		\begin{tabular}{c|ccc}
			 & $a$ & $b$ & $c$  \\
			\hline
			B\&G \cite{BauerPRD2023} & 0.4(1) & 2.1(2) & 0 \\
            Dual & 0.20(1) & 2.33(3) & -0.15(1)\\
			RDB \cite{Fontana2024} & 0.42(7) & 2.5(2) & -0.8(1)  \\
			Improved RDB & 0.28(6) & 3.2(2) & -1.4(1)  \\
		\end{tabular}
		\label{table2}
		\caption{Fit parameters for Eq. \eqref{log_fit_function} of the different computational bases considered in the main text, reported in the first column.}	
	\end{table}

    \section{\label{global_truncation} Global truncation}
    As discussed in the main text, aside of the local truncation of the gauge degrees of freedom to the first $L_{\text{max}}+1$ eigenstates of the local Hamiltonian, one can implement a global truncation restricting the total number of possible loop excitations by setting $\alpha_{\bf n_1}+\alpha_{\bf n_2}+\alpha_{\bf n_3} \leq N_{\text{max}}$. Combined with the parity symmetry, it can provide further improvement to the RDB. We show here further results on the performance of this global truncation.
    
    We compactify the labelling of the different truncations to
	\begin{equation}
		(L_{\text{max}})_{N_{\text{max}}}(\text{dim}(\mathcal{H}_{\text{phys}})),
		\label{truncation_label}
	\end{equation}
	where in the round brackets we explicitly write the Hilbert space dimension of the full lattice system. For example, the truncation label $6_8(86)$ indicates that when we retain locally $7$ states ($L_{\text{max}}=6$) and restrict to at most $8$ total loop excitations ($N_{\text{max}}=8$), the total size of the Hilbert space is reduced to $\text{dim}(\mathcal{H}_{\text{phys}})=86$. Without the combined parity symmetry and global truncations, the Hilbert space dimension for ($L_{\text{max}}=6$) would be $(L_{\text{max}}+1)^3=343$.

\begin{figure}[h]
		\centering
		\includegraphics[width=0.65\linewidth]{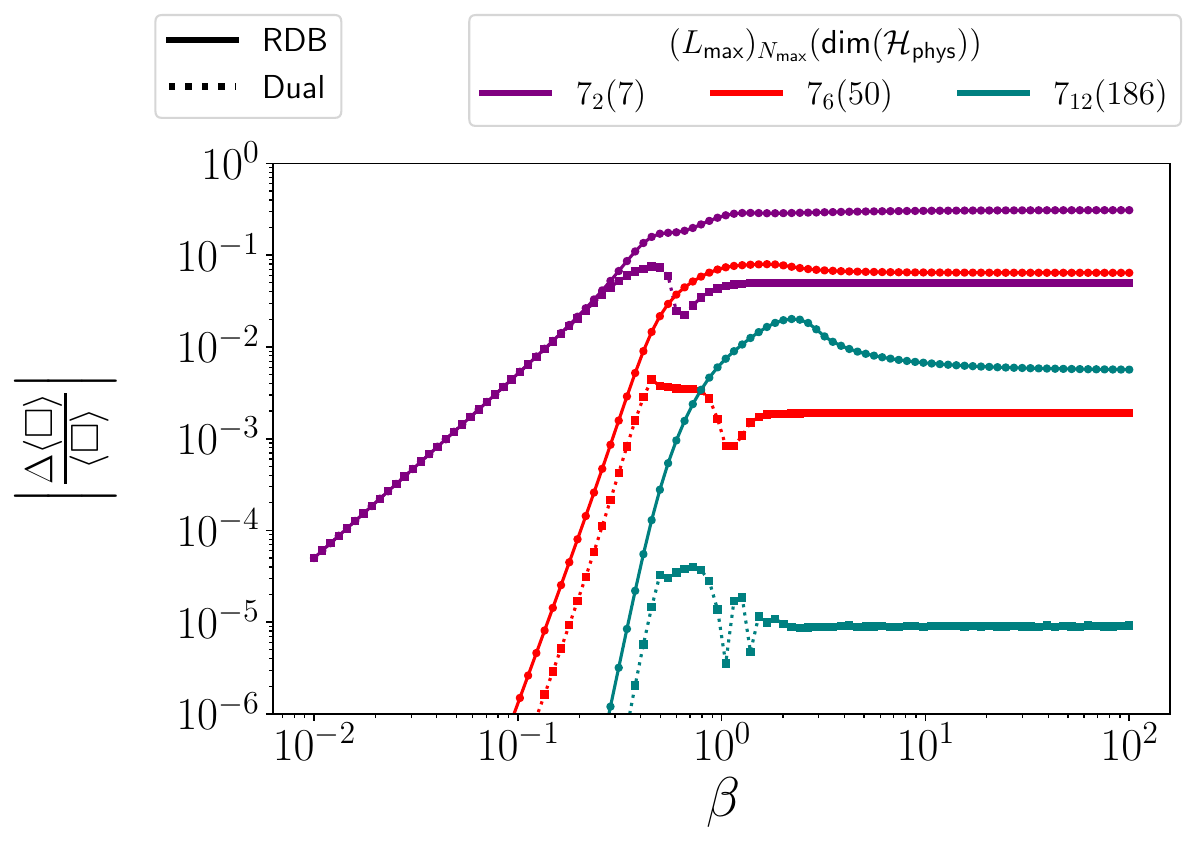}
		\caption{Relative precision on the plaquette expectation value for the 2+1D $\text{U}(1)$ LGT on the minimal torus as a function of the inverse coupling strength $\beta=1/2g^2$. The results correspond to a local truncation of the basis of $L_{\text{max}}=7$ and to varying global truncations of the number of loop excitations, $N_{\text{max}}$. Each data point has been obtained with (continuous lines, RDB) or without (dotted lines, Dual) variational optimization of the local basis.}
		\label{relative_plaq_parity}
	\end{figure}

    The global truncation can be implemented with or without the variational optimization of the basis. As an example, we show in Fig. \ref{relative_plaq_parity} the relative error in the computation of the plaquette observable with a local truncation of $L_{\text{max}}=7$ and different global truncations, with or without variational minimization. The reference values are those obtained with the local truncation of $L_{\text{max}}=7$ without any further truncation, with a total Hilbert space dimension of $\text{dim}(\mathcal{H}_{\text{phys}})=(7+1)^3=512$, optimized with the RDB. Fig. \ref{relative_plaq_parity} shows a significant reduction of the total Hilbert space dimension needed to reach a given precision, since the local truncation for each plaquette may still lead to highly excited states when combined with the other plaquettes. For example, the truncation  $(L_{\text{max}})_{N_{\text{max}}}=7_{6}$ with the RDB achieves at least per-cent level error while yielding a significant reduction of the Hilbert space. Only implementing exactly the parity symmetry reduces the Hilbert space dimension from 512 to 256.
    
	\section{\label{local_basis_computations}Computations for the Renormalized Dual Basis}
    As discussed in the text, the RDB is obtained through the resolution of the single-plaquette problem. Thanks to gauge invariance, the problem can be reduced to a single gauge degree of freedom. We restrict here to $U(1)$ LGT, even if the following points can be generalized to other groups \cite{Fontana2024}. 

    We parametrize $\text{U}(1)$ group elements as $g=e^{i\theta}$ with $\theta\in\left[-\pi,\pi\right)$. The group $\text{U}(1)$ is equipped with the Haar measure $\frac{d\theta}{2\pi}$, where the factor of $2\pi$ ensures that the function $f(\theta)=1$ is normalized to unity. The Hilbert space of a single gauge degree of freedom is the space of rays in $L^2(\text{U}(1))$, the set of square-integrable functions over the group. The set $\{|g\rangle\ |\; g\in \text{U}(1)\}$ forms a basis of $L^2(\text{U}(1))$, where $|g\rangle$ is the delta distribution, compatible with the Haar measure, peaked at $g\in \text{U}(1)$. 
    
    Since $\text{U}(1)$ is an Abelian group, all its irreducible representations (irreps) are one-dimensional. They are labelled by an integer $k\in\mathbb{Z}$, with the $k$-th irrep given by the function $f_k:\text{U}(1)\rightarrow\mathbb{C}$ defined as $f_k(e^{i\theta})=e^{ik\theta}$.
    According to the Peter-Weyl theorem, the functions $f_k$ define another basis for $L^2(\text{U}(1))$, which is a re-statement of the existence of the Fourier transform \cite{Knapp}. In the strong-coupling limit $g\rightarrow\infty$, the Hamiltonian reduces to the electric field term $E^2$, which corresponds to the quadratic Casimir in group-theoretic terms. In the parametrization above, $E^2$ turns into the differential operator $-\partial^2_\theta$ acting on $L^2(\text{U}(1))$. The functions $f_k$ are eigenfunctions of this operator, i.e., $E^2f_k = k^2f_k$. For this reason, the basis $\{f_k | k\in\mathbb{Z}\}$ is known as the electric basis.

    As presenteed in the main text, for general $g$, the LGT problem defined on a single square translates into the differential equation
	\begin{equation}
		H_{0}(g)=-2g^2\partial^2_{\theta}+\frac{1-\cos\theta}{g^2}.
        \label{singleHam}
	\end{equation}
    Note that the operator $T: L^2(\text{U}(1))\rightarrow L^2(\text{U}(1))$, defined by $T|e^{i\theta}\rangle=|e^{-i\theta}\rangle$, is a $\mathbb{Z}_2$ symmetry of the Hamiltonian \ref{singleHam}. Since $\left[H_0 (g),T\right]=0\; \forall g\in\mathbb{R^+}$, the energy eigenstates can be chosen to transform either in the fundamental or the sign representations of $\mathbb{Z}_2$. At strong coupling $g\rightarrow\infty$, the subspace spanned by the functions $f_k$ and $f_{-k}$ is degenerate for all $k\in\mathbb{Z}, k\neq0$. In this case, the $\mathbb{Z}_2$ symmetry can be diagonalized within the subspace, yielding the functions $\cos(k\theta)=\frac{1}{2}\left(e^{ik\theta}+e^{-ik\theta}\right)$ and $\sin(k\theta)=\frac{1}{2i}\left(e^{ik\theta}-e^{-ik\theta}\right)$.
    
    The eigenstates of the Hamiltonian can be obtained through numerical resolution of the differential equation \ref{singleHam}. However, it is more convenient to expand them in the electric basis. Given the $T$ symmetry, even (odd) functions can be written in terms of $\cos(k\theta)$ ($\sin(k\theta)$). Using the electric basis $f_k$ to truncate the full lattice system --either on a classical or quantum computation-- would require an impractically large number of states at weak coupling $g\rightarrow0$. However, for the computation of the RDB only a modest number of states is needed, well within the reach of any classical computer.
    
    To compute the RDB we fix $n_{\text{trunc}}\in\mathbb{N}$ (not to be confused with the global truncation $N_{\text{max}}$ described in the previous Appendix) and restrict to the subspace spanned by $\cos(k\theta)$ with $0\leq k\leq n_{\text{trunc}}$ for the even sector, and by $\sin(k\theta)$ with $1\leq k \leq n_{\text{trunc}}$ for the odd sector. In each subspace, the matrix elements of all relevant operators are obtained analytically. The operators $E^2$ and $\cos(\theta)$ already appear in the local Hamiltonian \ref{singleHam}, while $E$ and $\sin(\theta)$ enter in the full lattice system. Below we exemplify the required computations for the even sector; the odd sector is computed analogously. The properly normalized functions are $\sqrt{2}\cos(k\theta)$ for $k\neq0$ and $f(\theta)=1$ for $k=0$. We already know the operator $E^2$ is diagonal in the electric basis, while the matrix elements of the $\cos(\theta)$ operator are obtained analytically through the trigonometric relations
    \begin{equation}
        \cos(\theta)\cos(k\theta)=\frac{1}{2}\left[\cos((1+k)\theta)+\cos((k-1)\theta)\right],
    \end{equation}
    \begin{equation}
    \begin{split}
    \cos(k_1\theta)\cos(\theta)\cos(k_2\theta) 
    &= \frac{\cos(k_1\theta)}{2}\left[\cos((k_2-1)\theta)+\cos((k_2+1)\theta)\right] \\
    &= \frac{1}{4}\Big[ 
          \cos((k_1+k_2-1)\theta) 
        + \cos((k_1-k_2+1)\theta) \\
    &\hspace{2cm}
        + \cos((k_1+k_2+1)\theta) 
        + \cos((k_1-k_2-1)\theta) 
    \Big].
    \end{split}
    \end{equation}
    For $k_1,k_2\neq0$ the $\cos(\theta)$ matrix element is then
    \begin{equation}
    \begin{split}
        &\langle \sqrt2\cos(k_1\theta)|\cos(\theta)|\sqrt2\cos(k_2\theta)\rangle=2\int_{-\pi}^\pi\frac{d\theta}{2\pi}\cos(k_1\theta)\cos(\theta)\cos(k_2\theta)\\
        &=\frac{1}{2}\int_{-\pi}^\pi\frac{d\theta}{2\pi}\left[\cos((k_1+k_2-1)\theta)+\cos((-k_1+k_2-1)\theta)+\cos((k_1+k_2+1)\theta)+\cos((-k_1+k_2+1)\theta)\right]\\
        &= \frac{1}{2}(\delta_{k_1+k_2-1}+\delta_{-k_1+k_2-1}+\delta_{k_1+k_2+1}+\delta_{-k_1+k_2+1}) = \frac{1}{2}(\delta_{-k_1+k_2-1}+\delta_{-k_1+k_2+1})
    \end{split}
    \end{equation}
    where, in the last equality, the terms $\delta_{k_1+k_2-1}$ and $\delta_{k_1+k_2+1}$ have been removed as they are 0 for $k_1,k_2>0$. For $k_1=0$, the matrix element is
    \begin{equation}
    \begin{split}
        &\langle 1|\cos(\theta)|\sqrt2\cos(k\theta)\rangle = \frac{1}{\sqrt2}\int_{-\pi}^\pi\frac{d\theta}{2\pi}\left[\cos((k+1)\theta)+\cos((k-1)\theta)\right]=\frac{1}{\sqrt2}\delta_{k-1},
    \end{split}
    \end{equation}
    where in the last equality the term $\delta_{k+1}$ has not been included as it is also 0 for $k>0$.

    Analogous computations are carried out for the $\sin(\theta)$ operator and for the odd subspace. For fixed $g$ and truncation $n_{\text{trunc}}$, the lowest lying states of the single-plaquette problem \ref{singleHam} are obtained through exact diagonalization. Once these states are computed, the expectation values of the relevant operators ($E^2$, $E$, $\cos(\theta)$, $\sin(\theta)$) are calculated and stored. This procedure is repeated over a sufficiently dense grid of coupling strengths $g$, enabling the interpolation of all required matrix elements across the full range of couplings. The computation of the RDB and its matrix elements is a one-time precomputation. In subsequent computations, all the operators appearing in the full lattice Hamiltonian can be decomposed in terms of these precomputed ones, since they come from tensor product of the single-plaquette operators. 
	
	\bibliography{biblio}
	
\end{document}